\pdfoutput=1
\documentclass[conference]{IEEEtran}
\usepackage{url}
\usepackage{amsmath}
\usepackage[pdftex]{graphicx}
\usepackage{epstopdf}

\usepackage{caption}
\usepackage{subcaption} 

\usepackage[linesnumbered,ruled]{algorithm2e}

\DeclareMathOperator*{\argmax}{arg\,max}

\begin{document}

\title{Passive Fingerprinting of SCADA in Critical Infrastructure Network without Deep Packet Inspection}

\author{\IEEEauthorblockN{Sungho Jeon, Jeong-Han Yun, Seungoh Choi, Woo-Nyon Kim}
\IEEEauthorblockA{The Affiliated Institute of ETRI\\
Email: \{sdeva, dolgam, sochoi, wnkim\}@nsr.re.kr}
}
\maketitle




\begin{abstract}

We present the first technique of passive fingerprinting for Supervisory Control And Data Acquisition (SCADA) networks without Deep Packet Inspection (DPI) and  experience on real environment. Unlike existing work, our method does not rely on the functions of a specific product or DPI of the SCADA protocol. 
Our inference method, which is based on the intrinsic characteristics of SCADA, first identifies the network port used for the SCADA protocol, then consecutively infers the field devices and master server. We evaluated the effectiveness of our method using two network traces collected from a real environment for a month and a half, three days from different CI respectively. This confirmed the ability of our method to capture most of the SCADA with high F-score nearly 1, except for HMIs connected to master server, and demonstrated the practical applicability of the method.
\end{abstract}

\IEEEpeerreviewmaketitle

\section{Introduction}
\noindent\textbf{Motivation.}
Disaster caused to critical infrastructure (CI) by cybercriminals have been discovered since the damage resulting from Stuxnet and subsequent attacks \cite{farwell2011stuxnet}. 
Here it is noteworthy that cybercriminals' objectives for CI seem different from their intentions with ordinary IT network malware. In the case of CI, the attack is designed specifically for a certain operation and the attackers do not reuse the method after their attack is discovered. 
For this reason, not only inspection for real cite by security expert, we also need to introduce a security device well suited to CI such as anomaly-based IDS designed for Supervisory Control And Data Acquisition (SCADA) \cite{SRID_ESORICS14}, instead of fully depending on the introduction of a signature-based antivirus program.
  
  Unfortunately, in reality, a commonly occurring desperate situation is that the CI administrator does not have information about all the assets forming part of the CI or provides wrong information. This happens because the information is not updated, or because the maintenance is delegated to another IT company or the owner of the infrastructure is misinformed after maintenance by the vendor. It is a critical problem for applying mentioned security measures. 
  Furthermore, mapping in the absence of prior information is essential part of disaster recovery in critical infrastructure. Importance of this problem is emphasized in the recent DARPA call for proposals \cite{darpaCFP}. Therefore, there is a need for a system that provides information that would enable the target CI network to be understood for efficient site-inspection and to ensure that the administrator exactly understands how their network is configured.

\noindent\textbf{Limitation.} 
An active fingerprinting approach, which provides accurate results by interrupting the system, is not allowed in a real environment, because most of the devices used in CI are either out-dated or have low computational power; thus, the network could crash as a result of any small interruption. However, it may not feasible to use a existing passive fingerprinting study of a SCADA network in real environment. The limitation of existing work is that it relies on a specific environment, as an example, \cite{antfarm} required a parsing module of the target SCADA protocol to identify which device is related to the SCADA part. However, it may require extremely high cost because each CI typically consists of systems and products with different specifications deployed by different vendors. Furthermore, some vendors have designed a proprietary SCADA protocol, they decline to provide the customer with the protocol specifications; instead, they prefer to support visits by an expert or the use of a remote service not only to address security problems, but also to prevent technology leakages. The customized protocol from standard also cause intefere success of parsing. Therefore, existing work cannot cover all of real cases for parsing SCADA network trace to identify their network shapes. 
  
\noindent\textbf{Our Approach.}
In this work, we propose a novel passive fingerprinting method to identify the SCADA part in CI without deep packet inspection (DPI) to avoid dependence on a specific vendor. Specifically, our goal is to provide information as to which devices belong to the SCADA part, such as a field device and the master server by analyzing the network trace. 
Our approach is based on the intuition and observation that although each CI is associated with different vendors and products, typically SCADA has intrinsic characteristics in common with others in terms of their purpose and structure, despite their respective diverse customized environments.  

In practice, we collected two CI network traces consisting of about 2.9TB of data and 0.37TB in a real environment during normal operation. We represented this network trace in our segmentation definition. Using these pre-processed data, we first infer the network port used by the SCADA protocol (to which we refer as the SCADA port) based on our ranking function, instead of identifying SCADA protocol type. This port is identified based on our proposed algorithm with five features: \textit{Periodicity}, \textit{Communication Durability}, \textit{Device Complexity Gap}, \textit{Service Popularity}, and \textit{Segment Size}. Next, we inferred SCADA devices in the CI, including field and master server through common SCADA characteristics such as the connectivity degree.

The key advantages of our methodology are as follows: (1) Our passive fingerprinting approach is able to solve problems for which the administrator either does not have the exact information or has to verify their information. It is especially required when we have to take surveilance for  a plenty of our CI in the field not just a few target CI (2) Our approach does not leverage DPI, unlike existing work; hence, ours is a generic technique in terms of SCADA vendor and product.

\noindent\textbf{Contribution.}
The main contributions of this paper are summarized as follows: 
\begin{itemize}
	\item To the best of our knowledge, for the first time, a passive fingerprinting method is proposed to identify which devices on the CI network are SCADA devices without the need for DPI. Our method categorizes devices into three types as follows: (1) SCADA field devices, (2) SCADA master server, and (3) non-SCADA devices known as peripheral devices. We also showed that we can find HMI in three layer architecture when we have pre-knowledge that it consists of three layer.
	\item We present a experiene of handling real SCADA network trace with our data representation to leverage intrinsic characteristics of SCADA with verification and key observation of common knowledge of the SCADA network with about 2.9TB and 0.37TB of network traces collected in a real environment.
	\item Our method distinguishes most SCADA devices from other devices, except for the secondary HMI, and it shows that the analysis of a 6\% proportion of the dataset (which required about 2.5 days to collect) can produce the same result as the entire dataset with approximately 75 minutes of processing time in our dataset1.
	\item We evaluate generality of our method with two different types of network traces, SCADA consists of three layers with single SCADA protocol usage and consists of two layers with two proprietary SCADA protocol usage, respectively \footnote{Our experiment source code will be public via Github for reference without dataset, once accepted.}.
\end{itemize}

\noindent\textbf{Roadmap.}
We first provide an overview of existing work in Section 2. Next, we present the preliminaries including the target environment, threat model, data description, and data pre-processing in Section 3. In Section 4, we explain our methodology, and this is followed by the results we obtained with our method in Section 5. In Section 6, we discuss interesting questions related to our methodology. Finally, we conclude in Section 7.

\section{Related Work}
\noindent\textbf{Analysis for Critical Infrastructure Network.} 
 A few groups have reported work in connection with the analysis of control system networks. For example, a passive analysis tool capable of providing information about the network architecture has been described \cite{smartcisco}. Other workers developed a tool \cite{netapt} to review their security policy to check whether it is configured as intended based on firewall configuration data. However, the limitations of the aforementioned work are similar to those mentioned in Section 1 for \cite{antfarm}. Furthermore, these methods are unable to reveal details of devices identified in the SCADA network, such as whether they are field devices or master server. 
    
\noindent\textbf{Passive SCADA Device Fingerprinting.} Most SCADAs contain out-dated systems or devices which have been operating for a few years. These systems are vulnerable to a slight increase in overhead and have low computational power. These are common characteristics of the SCADA world, i.e., the long life cycle of their devices, that complicate the application of active fingerprinting techniques to real SCADA systems. The fact that the use of these techniques could possibly compromise the availability and safety \cite{moteff2004critical} of SCADA components is considered the reason why the passive approach for fingerprinting is highlighted in SCADA domains.
  
  The most common passive approach in device fingerprinting is to leverage TCP/IP information \cite{franccois2011ptf}. On the other hand, other studies of fingerprinting for SCADA devices that have recently been proposed are not based on network tracing, but rely on port scanning for specific SCADA protocols instead. A study aiming to identify field devices \cite{gordeychik2013scada} relied on specific SCADA protocol functions; for example, the Modbus protocol provides a function that responds with device information. The appearance of search engines to search for devices connected to the Internet \cite{matherly2009shodan} has motivated  research \cite{shine2014, kiravuo2015peeking} using methods other than finding SCADA devices using port scanning. However, port scanning, categorized as active fingerprinting, has the limitation in that it can interrupt or halt devices with low calculation power or those that are out-dated legacy devices. Therefore, active fingerprinting of SCADA networks is not allowed in the real environment to prevent loss of gain and life-threatening disaster. The problems associated with SCADA device fingerprinting were also addressed by \cite{bodenheim2014evaluation, caselli2013feasibility}, but they only explain existing work and do not suggest a specific method. In summary, to the best of our knowledge, the use of a network trace without DPI to identify devices in a SCADA network has not been studied to date. 

\noindent\textbf{Protocol Reversing}
  The other applicable existing work is protocol reverse engineering that the process of the inferring the specification of an unknown network protocol \cite{lin2008automatic}. These techniques leverage approach for automatically extracting information of unknown target protocol. Even though they contribute to analysis of unknown network environment and helpful to provide major SCADA protocols, it is difficult to cover for analysis of all our CIs in the field. The main difficulty of this problem is closed SCADA information that cannot be provided in public. These techniques based on assumption that we require enough samples from target environment or classification from other distinct types of protocols. Understandably, our CI network information is not public and some major vendors do not public their SCADA protocol specification and information. Even though, these techniques help to analysis several famous SCADA protocol, but this approach is extremely expensive cost to cover all of environment. As SCADA market grows, our CI environment becomes heterogeneous for SCADA device and network protocol. Considering the possibility that customized protocol from standard also can make different result from identified signatures, our method can be applied complementary for analysis of CI that cannot be covered by protocol reversing efforts.

\section{Preliminaries}

\subsection{Target System Environment and Assumption}

  
  We should note that, we do not aim to suggest a method that is completely generic in terms of all types of SCADA architectures; rather, our method is aimed at a common architecture consisting of two layers. Based on our experience, this work is conducted under the following assumptions and targets: 
  
\begin{itemize}
	\item We aim to find two layer SCADA architecture consists of field devices and their master server. Middle level layer is omitted in recent SCADA configurations due to improvement of field devices, unlike traditional three-layer architecture consists of HMI responsible for network management by operator, field devices such as PLC, and middle level layers used to reduce overhead of HMI and field device. However, we also provide clue for finding HMI when we already know that it is configured three-layer architecture.
	\item We assume that we would be unable to identify the deployed SCADA protocol type and that the network port used for their SCADA protocol is not revealed, since we have no pre-knowledge of the SCADA protocol or application.
	\item We assume that we have knowledge for number of deployed SCADA protocol. In practice, at least, SCADA operator typically knows name of their SCADA vendor or it can be identified by finding contract history.
	\item We assume that the design of the SCADA protocol is based on the TCP layer. Although some SCADA protocols have custom-designed architectures, SCADA protocols based on TCP are increasingly used for reasons of compatibility.
\end{itemize}
    

\subsection{Data Description and Pre-Processing}
\noindent\textbf{Data Description.} 
  We collected two network traffic packet traces in \textit{pcap} format from the national CI during normal routine operation. The names and sites of the CI cannot be revealed for security purposes and identity information is redacted in this paper. We used a network tapping device with Wireshark, for collection that would not disturb their normal routine. In the first dataset, the total size of raw pcap files is about 2.9TB during a period of a month and a half. The number of total raw packets is about 4.22 billion and DNP3 \cite{clarke2004practical} is applied for SCADA protocol, which uses 20000 ports to control field devices. In second dataset, 0.37TB total size of raw pcap file for 3 days with 0.12 billion and two proprietary SCADA protocol is deployed by different vendors. We believe that our target CI belongs to the category of complex CIs and that this dataset is sufficiently large to reflect the real world environment of CI, rather than merely a specific aspect.

\noindent\textbf{Data Filtering using TCP Information.}  
  We filter two types of packets. 
  We filtered packets which use several common network service ports under the assumption that these service ports will not be changed easily, because of compatibility with other systems. Second, we filtered packets unrelated to the type of packets suitable to the purpose of our analysis, such as ICMP packets that are used to notify network errors. In this stage, eleven common types of network service packets are filtered about 7\% of the original raw packets (4.22 billion $\rightarrow$ 4.05 billion packets) in our dataset1. 

  
\noindent\textbf{Communication Segmentation.}
  Our analysis of network traffic is based on communication segments containing groups of related packets, i.e., our method does not function at the packet level. There are three reasons why our approach differs from that used in existing work. First, grouping accompanying packets used for a certain command or service is an exact way of measuring periodicity, because a few packets sent are immediately accompanied by an acknowledgment in a communication. Second, we observed that SCADA communication characteristics influence the inter-arrival time analysis unlike ordinary TCP communication. For an concrete example, the pattern and length of DNP3 communication is varied according to the payload \cite{jeon2014obfuscation}. We observe that this is not own characteristics of DNP3, but common in SCADA protocol. Third, our dataset is an exception in that it contains data resulting from constant attempts to establish connections to specific non-responding systems. In our experience, it usually happens when a different vendor is involved in changing the old system without appropriate considering for existing system.

\begin{figure}[ht!]
  	\centering
  	\includegraphics[scale=0.35]{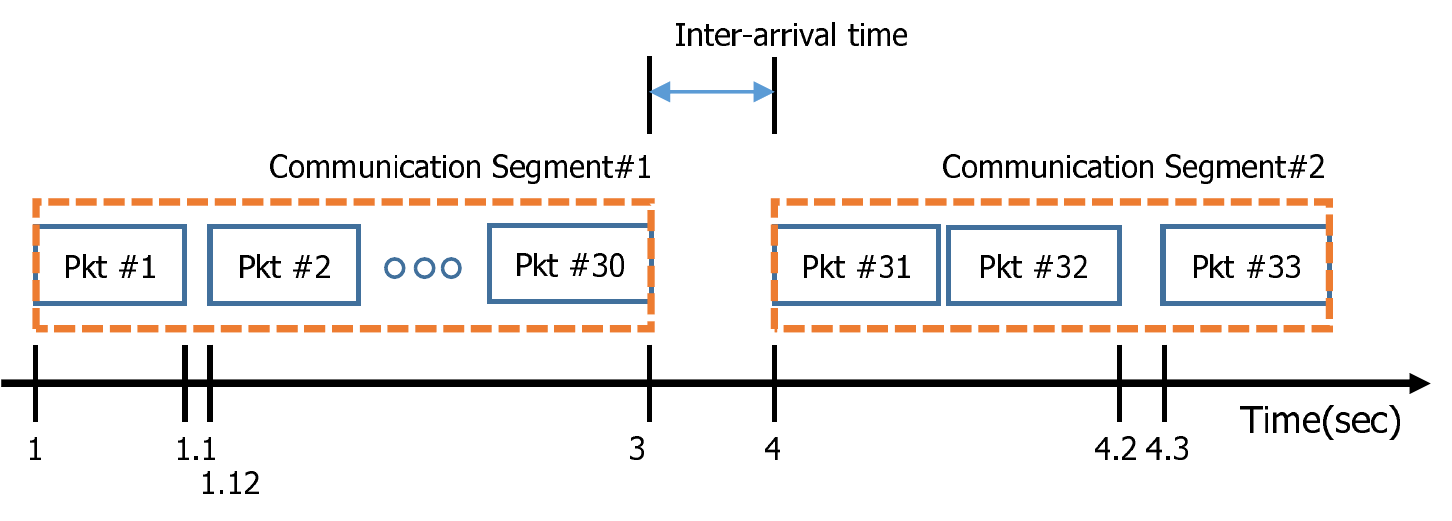}
  	\caption{Example of communication segment (\textit{$t_{comm}$}=1 second)}
  	\label{CC_figure}
  \end{figure}

  Considering these problems, we represent a network trace as a communication segment for which the next packet has not arrived within \textit{$t_{comm}$} threshold period for communication separation (Figure \ref{CC_figure}). 
  We heuristically set \textit{$t_{comm}$} as a second because the periodic monitoring command has a periodicity that is higher than at least a second so as not to cause severe overhead to the field devices. For instance, the WirelessHart sensor from Emerson can configure the sensor device to report every second as maximum frequent reporting, but in reality, it is configured to more than that due to battery problems. In addition, the excessive overhead caused by unnecessary frequent reporting can increase the maintenance cost. After processing the data with our communication segmentation policy, we aggregated data by reducing the data by about 98\% for the analysis of filtered packets (4.05 billion filtered packets $\rightarrow$ 2.06 billion communication segments).
  
  
  \vspace{-0.1in}
    
\noindent\textbf{Data representation: SCADA Segment 5-tuple. } \label{5-tuple}
  We analyzed the network trace based on our representation that the SCADA segment 5-tuple (\textit{ft}) data consist of both the IP addresses and port numbers of the source and destination ports, and the segment size (Equation. \ref{equ:ft}) with their time information.
  \begin{equation}
  \label{equ:ft}
  	ft := <SrcIP, SrcPort, DstIP, DstPort, SegSize>
  \end{equation}
   Note that the segment size is defined as the sum of the size of packets belonging to the same segment where the unit is a byte. Our dataset1 contained about 22.5 million different types of \textit{ft} with 366 device IP addresses from 2 billion communication segments. The large number of 5-tuples with the small number of hosts indicates that there are services based on a dynamic port policy, which uses a different port for each connection, and that a range of diverse services are used in our target CI. 
  

    
\section{Methodology}

\subsection{Overview}
  Our goal is to create an inside network map of the target CI by only using the network trace as input. For the sake of our goal, (1) we first make ranking list for SCADA communication based on our proposed features (2) we infer the network port used for the SCADA protocol from the ranking list, then (3) we infer SCADA devices such as field devices, the master server from SCADA. If we have one more than deloyed SCADA protocol, then we remove identified SCADA information from identified port, then repeat process from inferring SCADA port. We note that our method require additional information that the number of deployed SCADA protocol that can be provided by administrator or survey of contractor. 
  

\begin{algorithm}
	\SetKwInOut{Input}{Input}
    \SetKwInOut{Output}{Output}
	
	\Input{CI network trace dataset}
	\Output{SCADA device lists}

	Pre-processing for network trace\;
	Make SCADA Communication ranking list\;
	\For{i=0 : num\_SCADA\_Protocol}
	{
		Infer SCADA port from top ranked communication\;
		Find SCADA field device and master server using inferred SCADA port\;
		Remove inferred SCADA communication from ranked list\;
	}
	\caption{Infer SCADA device list Algorithm}
\end{algorithm}

  
\subsection{SCADA Protocol Port Inference} \label{SCADAPortInfer}
\noindent\textbf{Observation and Intuition. } \label{Obs} 
    Our approach toward SCADA protocol port inference is based on our intuitions and observations from experience. First, the network port used in field devices for connection with the master server is fixed at the same SCADA site. 
    Second, SCADA protocol connections have the tendency to maintain their connection longer than other forms of communication \cite{caselli2013feasibility}. 
    Third, we observed that about 99.4\% of SCADA communication in our dataset originates from monitoring field devices, i.e., this communication is periodically scheduled and automatically operated based on an event-polling approach. We should note that we observed that DNP3 communication has obvious periodic characteristics only when considering each pair consisting of a function code and object. For a concrete example, in our dataset, most of the ``Read-Response'' DNP3 communication was sent by $object_{A}$ (77.74\% of the total object ratio in ``Read'' function code) whose inter-arrival packet distribution has a mean and variance of 8.75 and 1.48 seconds, respectively. The proportion of each object varied as a relation of the master server-field device. This result reflects the complexity of a real environment in which the CI consists of many field devices configured for a different role and purpose, and not using identical settings despite usage of a single SCADA protocol. In summary, even though SCADA protocols are different and have an unreadable format, periodicity is an intrinsic characteristic of SCADA communication, but we need appropriate method to utilize this characteristic. 




    
\noindent\textbf{SCADA Port Infernce.} 
    We first aim to infer the SCADA port based on a passive approach without DPI by considering this problem as a ranking problem with proposed function. 
    We first identify communication corresponding to the relationship between the master server and field device from ranking list. Then, we infer the SCADA port as the port used in the field device in the inferred SCADA communication. Our ranking function comprises five features: \textit{Periodicity} (pR), \textit{Communication Durability} (dR), \textit{Device Complexity Gap} (cR), \textit{Service Popularity} (uR), and \textit{Segment Size} (sR).   
    
    Specifically, we rank given a dataset $FT=[ft_1,...,ft_n]$ according to our ranking function, where $ft_i$ is  communication based on our \textit{ft} representation. Let $f$ denote a ranking function, which assigns a ranking value $f_i$ to each SCADA segment 5-tuple $ft_i$. Let $y=[y_1,...,y_n]$ denote a ranking score vector, where y is scaled to have a value from 0 to 1. This vector $y$ is computed by solving the following optimization problem:

    \vspace{-0.1in}

    \begin{equation}
      f^*=\argmax_{ft} pR(ft_i) * dR(ft_i) * cR(ft_i) * uR(ft_i) * sR(ft_i) 
    \end{equation}
    
\noindent\textbf{Feature1: Periodicity.} High \textit{periodicity} is the most well-known characteristic of SCADA communication. We measure the \textit{periodicity} of the inter-arrival time between segments in a \textit{ft} by defining periodicity as \textit{pRatio} and as follows: 
    \begin{equation}
      pR(ft_i) = \frac{mean_{IAT_{ft_i}}}{variance_{IAT_{ft_i}}}
    \label{equ_pRatio}
    \end{equation}
    where $IAT$ is the set of $iat$ that sums the inter-arrival time between communication segments corresponding to $ft_i$.

    Interestingly, unlike our expectation, SCADA communication is neither the most periodic communication nor the most highly ranked in both of our dataset. In our dataset1, there are 11 types of network services with 345 communications that are ranked higher than the most highly ranked SCADA communication with respect to \textit{pR} with the highest $pR$ being 23,862.69. These services are the heartbeat of network devices and printers, relationship with the backup and DB server, peripheral services such as Netbios, and Network Time Protocol (NTP). They are also configured to carry out scheduled tasks and their scheduled time for communication varied from 2 to 60 seconds. We should note this order of ranking does not indicate that SCADA communication is not periodic. It has sufficient periodicity, but is lower ranked because of a difference in variance of the order of milliseconds. 
                
    Another interesting insight is that the introduction of the segment size feature to data representation is crucial to capture the periodicity of SCADA communication. 
    In our dataset, none of the SCADA communication represented by a 4-tuple (from which the size feature from our \textit{ft} definition (Equation. \ref{equ:ft}) has been excluded) exceeds a p-ratio value of 1, which is indicative of un-periodic communication because of diverse configuration of real environment for each connection. 
    In our dataset, our data representation captures 557 (0.28\% in SCADA communications) periodic SCADA communications for $pR > 3$ and the highest $pR$ is 377.54. Definitely, the main reason why other cases that unsuccessfully capture periodicity are limited is that they cannot handle SCADA communication with the same size and pattern, but a different payload with periodic time. However, there are some cases with a different communication pattern and payload that can be exactly distinguished by segment size. 
    
    
\noindent\textbf{Feature2: Communication Durability.} It is well-known that SCADA communication tends to be maintained once established, because the purpose and usage of the communication is fixed when these aspects are configured. 
We exploited this characteristic by considering two statistics conjointly: the observed length of communication and the occurrence of corresponding communication. We named this feature \textit{Communication Durability} and defined it as follows:
\begin{equation}
\label{equ_lRatio}
	dR(ft_i) = \sum{IAT_{ft_i}} * \log{n_{ft_i}}
\end{equation} 
	where $n_{ft_i}$ is the occurrence of corresponding $ft_i$ in the total network trace. The unit of a feature is also important and we empirically decided to use hours for the observed length considering the relationship between the $\log$ of occurrence scores.


It should be noted that to decide the filtering threshold heuristically is difficult. For a concrete example, in our dataset1, although most of the communication segments have low occurrence values ($P(X\leq9)=0.9$, $P(X\leq324)=0.99$) 
in our dataset, it does not indicate which filtering threshold is desirable, which therefore completely depends on users' intuition. There are over five thousand types of occurrence value with the highest value being 72,921,879. Instead of heuristic threshold, our feature reflect this characteristics to scroe for ranking.  

  According to our second feature $dR$ in dataset1, the highest $dR$ communication is generated by the X11 service ($dR$=703.01) and the other highly ranked services are also related to remote services and network devices. Network devices do not have to be reset except when they experience severe problems and do not have to be updated when they are reset; thus, they have a lengthy relationship based on communication. The highest SCADA communication is ranked 258 ($dR$=368.04), proving that SCADA communication is also maintained for a long period. Even though field devices maintain their relationship for a long time, they sometimes require a reset for maintenance purposes or in the case of a mechanical problem. 

         
\noindent\textbf{Feature3: Device Complexity Gap.} In SCADA, there is a huge difference in terms of diverse network port usage between field devices and master server. This is because field devices only provide a simple function that involves periodic reporting and device status control in an emergency by fixing the SCADA port, but a master server has to communicate with many field devices simultaneously and provide support for other services. Our rough guess is that this difference in complexity according to the number of distinct network ports (i.e., the cardinality of the set of network ports) used by the devices is as follows:
\begin{equation}
\label{equ_complexity}
	cR(ft_i) = \frac{|Pt_{SrcIP}|}{|Pt_{DstIP}|},
\end{equation} 
where $Pt$ is the set of network ports used by $SrcIP$ in the total network trace. 

In our dataset1, all field devices only use a single network port for SCADA communication. In contrast, a master server is related to 1,147,082 \textit{ft} communications (5.10\% of the total number of \textit{ft}). It means that there is much difference between the usage of a field device and a master server in terms of network communication. The highly ranked $cR$ communication is not SCADA communication, but that of a peripheral device connected to HMI with the highest $cR$ of 63,350. These peripheral devices are also only configured to perform a na\"ive function such as backup or reporting. Although SCADA communications are not the highest ranked, they are highly ranked with the highest $cR$ as 32,918 (ranked from 29th to 75th out of a total of 742). We conclude that a small difference in the order does not have significant meaning, because it can fluctuate due to dynamic port allocation.



\noindent\textbf{Feature4: Network Service Popularity.} The purpose of this feature is to distinguish the relationship between a server and simple peripheral device from SCADA communication. The number of 3-tuples, which consist of $<Port, SrcIP, DstIP>$, denoted $PU_{(port,SrcIP,DstIP)}$, represent the degree of usage of this network port by diverse devices, and we refer to this property as service popularity (PU). A field device has an extremely high PU value, but a master server has a low value. This is because every field device uses the same SCADA port, but a master server uses dynamic port allocation. To exploit this idea, we define \textit{Network Service Popularity} ($uR$) as follows:
	\begin{equation}
	  uR(ft_i) = \frac{|PU_{(SrcPort, SrcIP, DstIP)}|}{|PU_{(DstPort, SrcIP, DstIP)}|}
	\end{equation} 
	or the inverse of $uR$ when $uR$ is less than 1. According to this feature, in our dataset1, the most highly ranked communication is a VPN service used for connecting an HMI to a remote sub-station and this is followed by SCADA communication.

  We note that the occurrence of a port itself cannot be a useful feature to identify SCADA communication. Because, in terms of occurrence, communication with lower periodicity naturally has higher occurrence compared with higher periodicity. In our dataset, communication for checking a device or service status are sent every a few seconds, but SCADA communications are scheduled to be sent every minute or about 8 seconds, typically in our dataset. We therefore do not consider occurrence of communication as a main feature.

\noindent\textbf{Feature5: Segment Size.}
  Main task of field device in SCADA is to report their measurement to master server to check operation status. Naturally, communication for reporting measurement has bigger segment size than others. Depends on environment, it is configured based on fragmentation packets, thus analysis based on segment is effective to capture these communications successfully. Typically, segment used for service status check has small segments size less than 100 bytes, but most reporting communications relatively higher segments size over 400 bytes. This feature is useful to discriminate from service status check communications used in master server.

        \begin{table*}[t]
    \caption{Top-5 \textit{ft} with respect to proposed ranking function (\textit{f}) in dataset1}
    \label{tbl:ranking_d1}
    \centering
    	\begin{tabular}{|c c c c c c | c c c c c | c|}
    		\hline
    		Rank & S-IP & S-Port & D-IP & D-Port & SegSize & $pR$ & $dR$ & $cR$ & $uR$ & $sR$ & \textit{f} \\ \hline
    		1 & FD.31 & 20000 & M & 51382 & 340 & 0.3801 & 0.4200 & 0.5175 & 0.3636 & 0.2246 & $6.7470\times10^{-3}$  \\
    		2 & FD.12 & 20000 & M & 34362 & 337 & 0.3198 & 0.4582 & 0.5175 & 0.3636 & 0.2226 & $6.1358\times10^{-3}$  \\
    		3 & FD.03 & 20000 & M & 55886 & 332 & 0.3117 & 0.4659 & 0.5175 & 0.3636 & 0.2193 & $5.9924\times10^{-3}$\\
    		4 & FD.03 & 20000 & M & 55886 & 296 & 0.4613 & 0.3320 & 0.5175 & 0.3636 & 0.1955 & $5.6342\times10^{-3}$\\
    		5 & FD.31 & 20000 & M & 51382 & 225 & 0.4070 & 0.4200 & 0.5175 & 0.3636 & 0.1486 & $4.7806\times10^{-3}$\\
    		\hline
    	\end{tabular}
    \end{table*}

            \begin{table*}[t]
    \caption{Top-5 and 13th \textit{ft} with respect to proposed ranking function (\textit{f}) in dataset2}
    \label{tbl:ranking_d2}
    \centering
    	\begin{tabular}{|c c c c c c | c c c c c | c|}
    		\hline
    		Rank & S-IP & S-Port & D-IP & D-Port & SegSize & $pR$ & $dR$ & $cR$ & $uR$ & $sR$ & \textit{f} \\ \hline
    		1 & FD.A1 & $Port_A$ & M & 58292 & 686 & 0.5872 & 0.6573 & 1.0000 & 0.2917 & 0.4531 & $5.1003\times10^{-2}$  \\
    		2 & FD.A1 & $Port_A$ & M & 61082 & 686 & 0.5872 & 0.3403 & 1.0000 & 0.2917 & 0.4531 & $2.6417\times10^{-2}$  \\
    		3 & FD.A1 & $Port_A$ & M & 58292 & 288 & 0.5872 & 0.6573 & 1.0000 & 0.2917 & 0.1902 & $2.1412\times10^{-2}$\\
    		4 & FD.A1 & $Port_A$ & M & 49179 & 686 & 0.6158 & 0.9999 & 0.2604 & 0.2500 & 0.4531 & $1.8181\times10^{-2}$\\
    		5 & FD.A4 & $Port_A$ & M & 58291 & 1086 & 0.4557 & 0.6502 & 0.2500 & 0.2917 & 0.7173 & $1.5498\times10^{-2}$\\
    		13 & M & 49170 & FD.B1 & $Port_B$ & 1514 & 0.3117 & 0.7441 & 0.1847 & 0.1667 & 1.0000 & $0.7140\times10^{-2}$\\
    		\hline
    	\end{tabular}
    \end{table*}

\subsection{SCADA Device Inference}
  Next, we infer the identity of each system device from the network traffic based on statistics of the \textit{ft} and our intuition with the inferred SCADA port. We infer the identity of devices from (1) field device, then (2) master server. 
    
\noindent{\textbf{Inferring field device.}} Our idea for the inference of a field device is based on the intuition that field devices communicate with a few devices, such as master server, only through the SCADA port for reporting purposes. To exploit this idea, we infer field device if device satisfied two conditions, propotion of SCADA communication and small connectivity degree. We first investigate the occurrence proportion of network port usage in each device, before selecting the set of devices that engage in SCADA communication. Among these devices, we only consider device as field device candidate, if the proportion of SCADA communication in a device accounts for more than half of their total communication. Next, among these field device candidates, we infer field device when it has smaller connectivity degree than threshold (we propose threshold as five from our experience). If it satisfies only first condition, it can be inferred as master server in our master server inference stage later.

Our first dataset contains a total of 52 devices that use SCADA communications, including 1 master server, 49 field devices, and 2 devices used for reporting. These field devices are only connected to a master layer. We observe that all 49 field devices only use SCADA communication for reporting and are only connected to a master server through the 20000 (DNP3) port. 
On the other hand, a master server has a higher connectivity degree (64 $>$ 1) and more \textit{ft}s (110,674 $>$ 53) than a field device. Our idea allows us to use the inferred SCADA port to easily identify field devices. From our algorithm, we can exlucde two non-SCADA devices which has SCADA communication, but introduced for statistics reportinng.


\noindent{\textbf{Inferring master server.}} A CI administrator can deploy one or more master server for their SCADA for each part respectively, but we cannot know their exact number. Without pre-knowledge of this information, we guess the set of master server when a device satisfies these condition: (1) they are connected to an inferred field device, (2) they engage in SCADA communication. 
Hence, all 49 field devices are connected to the server in dataset1 and 22 field devices from SCADA vendor A and 4 field devces from SCADA vendor B are connected to master server in dataset2.



\section{Results}

We evaluate our methodology by answering the following questions: (Q1) How well is our method able to identify SCADA of CI? (Q2) What should be the size of the network trace to expect a result? 


\subsection{Identified SCADA in Critical Infrastructure}

\noindent\textbf{Finding SCADA port from SCADA Communication Ranking.}    
  As our proposed features, we rank communications to find SCADA port in both dataset respectively (Table \ref{tbl:ranking_d1}, \ref{tbl:ranking_d2}). Note that we normalized each feature by their maximum value.
  We first select the top ranked \textit{ft} as SCADA communication representing the relationship between a master server and a field device through a SCADA protocol. After selecting the SCADA communication, the port used in a device with a lower connectivity degree is inferred as the SCADA port. In our dataset1, 953 of top-1000 communications have network port 20000 which indicates DNP3 communication and in top-100 in second dataset, 62 communications are corresponding to proprietary SCADA protocol A and 14 communications are SCADA protocol B. We observe that fixed network port usage on field device side in both dataset. In first dataset, DNP3 network port 20000 is inferred from top ranked communication (Table \ref{tbl:ranking_d1}). Same process is conducted in second dataset, but to find two deployed proprietary SCADA port, we first inferred first SCADA protocol network port, then find second one after removing identifieid first SCADA communication (network port is redacted in second dataset to hide their proprietary protocol). 

      \begin{figure*}[ht!]
  	\centering
  		\begin{subfigure}[b]{0.35\textwidth}
  			\includegraphics[scale=0.6]{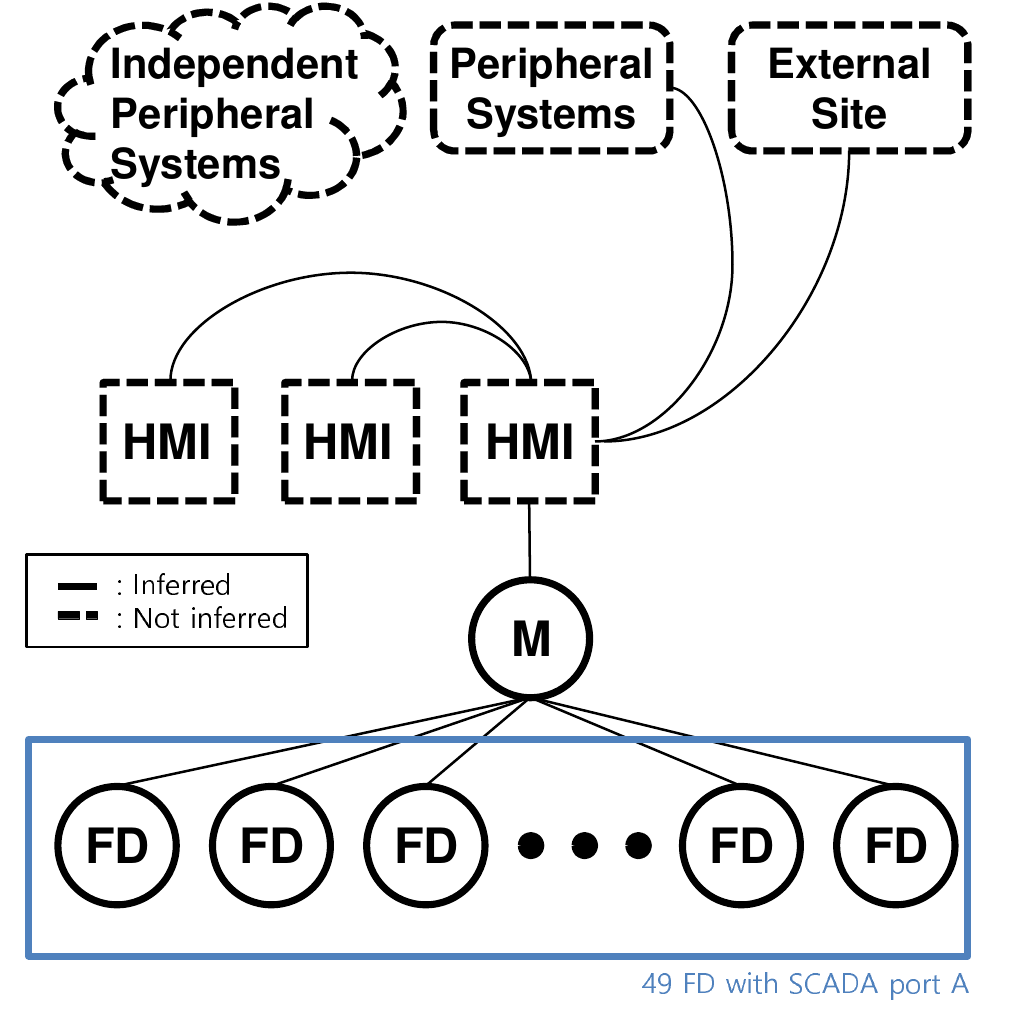}
  			\caption{In dataset1}
  		\end{subfigure}
  		\begin{subfigure}[b]{0.35\textwidth}
  			\includegraphics[scale=0.6]{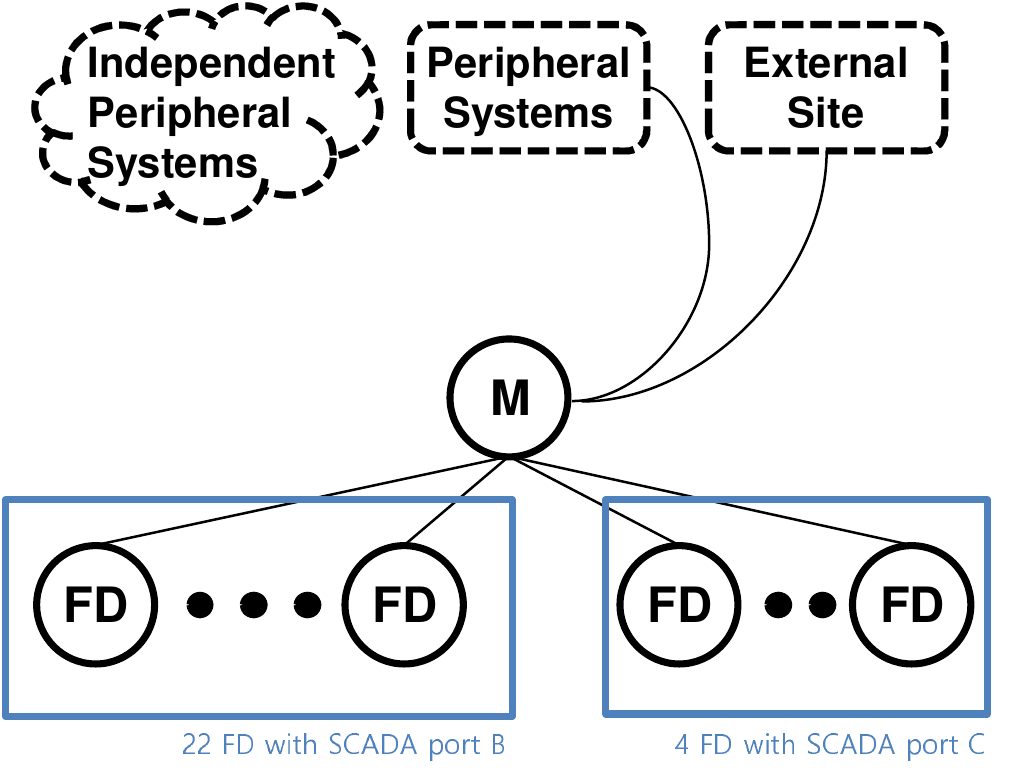}
  			\caption{In dataset2}
  		\end{subfigure}
  	\caption{Our target Critial Infrasturcutre network with inferred by our method (FD: field device and M: master server)}
  	\label{fig:infNet}
  \end{figure*}

\noindent\textbf{Finding SCADA devices using SCADA port.}
  Our method identifies all SCADA devices as two types, field devices and the master server, except for the HMIs introduced for monitoring in first dataset (Figure \ref{fig:infNet}). We provide method when we have preknowledge that HMI exist (Section \ref{subSec:inferHMI}). Once we identify a SCADA device, then we can try applying existing active-fingerprinting methods to the other device such that it will not cause catastrophic disaster even if it were to be interrupted unexpectedly.
   
  Next, we experiment with our second dataset to prove generality our our method. Interestingly, our second dataset deploys the two proprietary SCADA protocols at the same time for other purpose, thus task is to infer two cluster of SCADA devices respectively. Note that, existing work cannot be applied to this dataset because of their proporietary protocols. Our second dataset is collected from 2-layer SCADA architecture because it has modern SCADA field device that does not require middle layer device. Note that, this site has gateway device to deploy two SCADA protocol simultaneously. We regard this device as field device, because it is located in front of field device and only perform converting communication to provide compatibility. It identified 22 field devices for first SCADA group and 4 field devices for second SCADA group, and one master server 

  \noindent\textbf{Evalualtion measure: F-score.}
  Finally, we evaluate our method in terms of F-score with precision and recall. In both datasets, our method achieve high F-score from our method (F-score$_{d1}=0.9709$, F-score$_{d2}=1$). In our dataset1, our method successfully identified all field devices and master server (precision$_{d1}=1$), however, our method cannot automatically identify HMI deployed in dataset1 (recall$_{d1}=0.9434$). On the other hand, our method completely identified devices related two deployed SCADA vendors.
 
  \begin{figure}[ht!]
  	\centering
  	\includegraphics[scale=0.3]{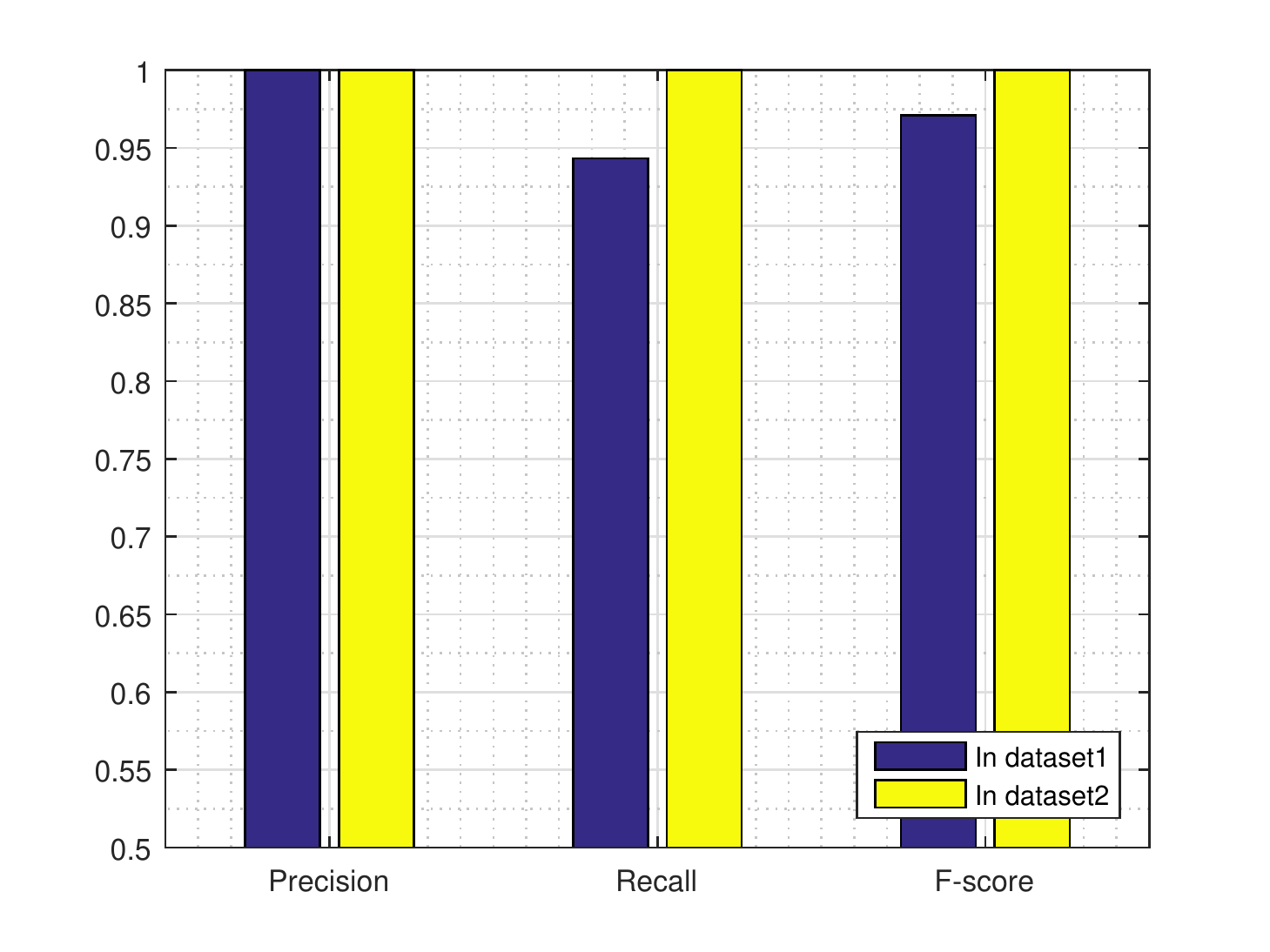}
  	\caption{Precision, reccall, and F-score from both datasets}
  	\label{fig:eval_fscore}
  \end{figure}

  \vspace{-0.1in}

\subsection{Traffic Volume Decision}
  The size of the network trace required for our method to be used successfully is an important question. We investigate the minimum size of the trace that would provide the same result as the total trace. In our dataset, about 6\% of the network trace corresponding to amount of 3 days is sufficient to produce almost the same result as the total dataset with 75 minutes of processing time. Most of the devices are discovered after collecting the trace for several minutes only, because, most of the devices are scheduled for less than a minute. However, communications for network devices are ranked higher than SCADA communication by our ranking function, when we experiment with less than 6\% of the network trace. We guess that the reason is a lower \textit{periodicity} value even though our target has higher values with respect to feature3 and feature4. It is not only caused by reason from margin caused by the network environment (Section \ref{SCADAPortInfer}), but also the limitation of capturing the exact SCADA communication type without DPI. 

\begin{figure}[ht!]
  	\centering
  	\includegraphics[scale=0.2]{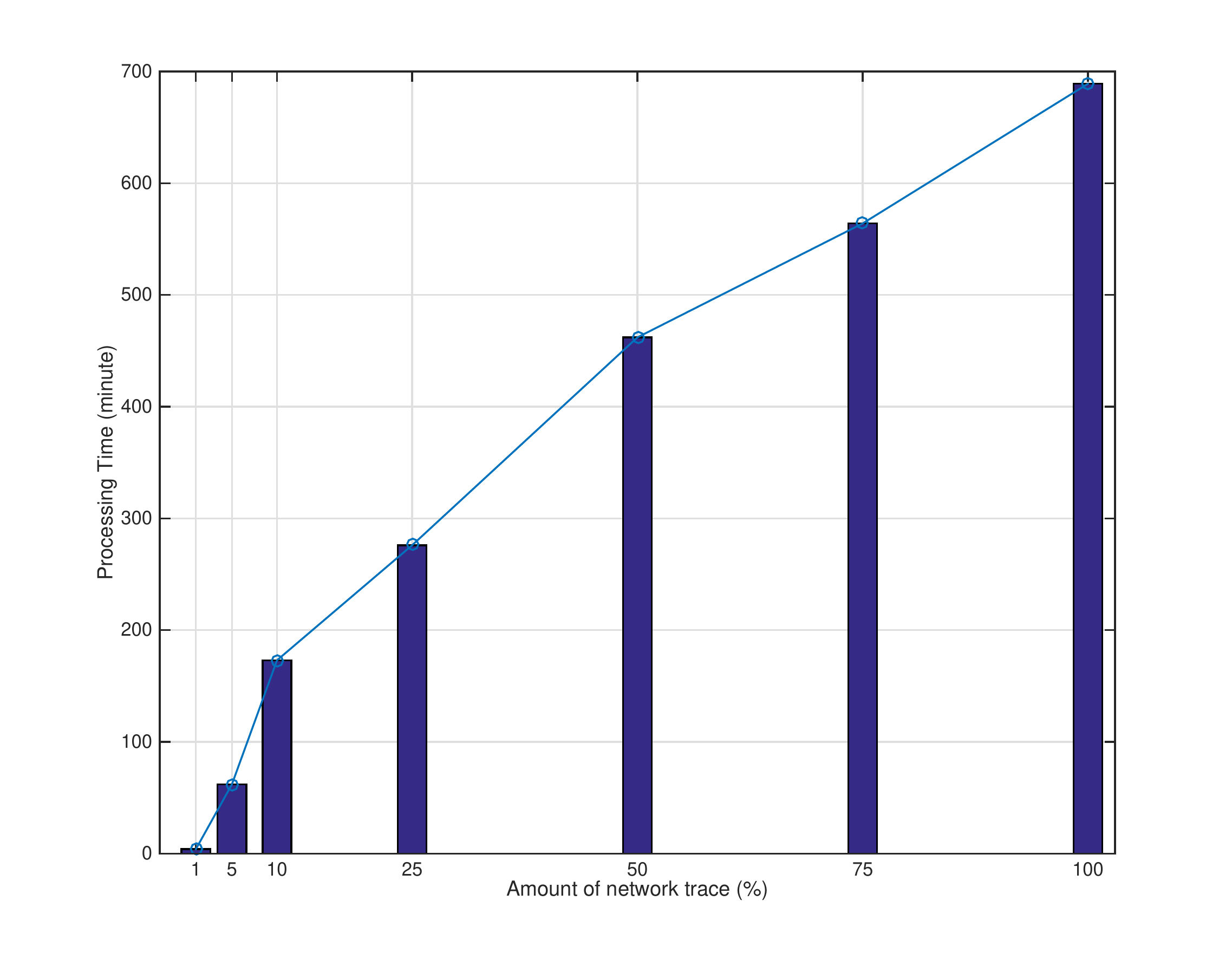}
  	\caption{Processing time (minutes) as a function of the size of the network trace (\%) }
  	\label{fig:pTime}
  \end{figure}

  
  We measured the processing time of our method as a function of the size of the trace. We measured on a system with an Intel Zeon Processor 2.3 GHz, 128 GB memory, SSD storage, and implemented by using Java 8 (Figure \ref{fig:pTime}). When the entire dataset was used for the experiment, it took nearly 11 hours to complete the computation. However, when we included less than 10\% of the dataset, it required less than 2 hours. 

  \vspace{-0.15in}
    
\section{Discussion}
  In this work, we only focused on identifying the field device and master server in SCADA part in the whole CI network, but we provide method to identify HMI when we have pre-knowledge that HMI is deployed in target CI. We also discuss \textit{Moving Target} research that can be counter-measure pontetially. 

\subsection {Inferring HMI with preknowledge} \label{subSec:inferHMI}
Additionaly, if we have pre-knowledge that target SCADA consists of three layer architecture, then we can infer HMI from inferred master server. We infer devices to which the inferred master server sends the largest amount of communication as HMI. The primary reason for communication between the master server and HMI is to deliver measurements collected from field devices. 
In detail, we define \textit{$Qty_{ft_i}$}, i.e., the quantity of communication, as follows:
  \begin{equation}
  	  Qty_{ft_i} = n_{ft_i} * SegSize_{ft_i}
  \end{equation}
  for a given \textit{ft}. Because, when we only consider the number of times communication is sent, it will be dominated by frequent types of communication such as network status or service check messages. However, $SegSize$ for measurement reporting is larger than for these types of communication; thus, we consider both factors. Then, we infer HMI as destination device in those relationships with the highest \textit{$Qty_{ft_i}$}, where SrcIP is the inferred master server and DstIP is one of the devices connected with the inferred master server.
  
  In our dataset1 whose SCADA consists of three layer architecture, we successfully capture the HMI through the relationship with the highest \textit{$Qty_{ft_i}$}, where SrcIP is the inferred master server with $7.55 \times 10^{13}$ of \textit{$Qty_{ft_i}$}. The second highest device is other peripheral devices for backup with $5.12 \times 10^{12}$ and the subsequent ranked devices are field devices with $1.76 \times 10^{12}$, $ 1.23 \times 10^{12}$, and $7.68 \times 10^{11}$, respectively. We should note that a limitation of our method in terms of HMI inference is that it cannot identify when multiple HMIs are applied. When there are many field devices that are difficult to monitor by a single HMI, multiple HMIs can be applied to monitor these field devices simultaneously. 

\subsection{Obfuscation and Moving Target}
Because of out-dated and low performance SCADA devices, it is difficult to apply encryption to existing SCADA. Even though it may have been studied for next generation of SCADA, introduction of a high-performance device causes a high expense for device requirements. A method that dynamically changes the properties of a system environment, called Moving Target Defense (MTD) or Randomization defense, can be a solution to this problem \cite{okhravi2014finding}. In recent times, it has been highlighted for diverse areas to reduce the attack surface and was also discussed as an introduction for SCADA \cite{ali2015randomization}. A sophisticated obfuscation study may become a countermeasure of our approach. Unfortunately, in our opinion, these defense techniques may be difficult to apply to a real environment in the near future.

\section{Conclusion}
  In this work, we present expereince of understanding of real environemt CI effectively. We propose a novel technique involving passive fingerprinting of SCADA in a CI network without the use of DPI, unlike existing work that either relies on a specific device function or on DPI to perform fingerprinting. 
Our method is based on a passive approach that does not influence the operating system, but performs an analysis based on a collected network trace by using techniques such as network tapping or network switch mirroring. Our work can be used for diverse security applications. Anomaly-based IDS can use our work to create an initial rule to define their normal behavior and the identity of the device. Our work is also expected to be useful for security experts sent to conduct a manual inspection of CI, or especially a CI administrator who does not have exact information of the assets under their management to prepare recovery.  


\bibliographystyle{IEEEtran}

\bibliography{./PASIN_DSN}

\begin{thebibliography}{10}
\providecommand{\url}[1]{#1}
\csname url@samestyle\endcsname
\providecommand{\newblock}{\relax}
\providecommand{\bibinfo}[2]{#2}
\providecommand{\BIBentrySTDinterwordspacing}{\spaceskip=0pt\relax}
\providecommand{\BIBentryALTinterwordstretchfactor}{4}
\providecommand{\BIBentryALTinterwordspacing}{\spaceskip=\fontdimen2\font plus
\BIBentryALTinterwordstretchfactor\fontdimen3\font minus
  \fontdimen4\font\relax}
\providecommand{\BIBforeignlanguage}[2]{{%
\expandafter\ifx\csname l@#1\endcsname\relax
\typeout{** WARNING: IEEEtran.bst: No hyphenation pattern has been}%
\typeout{** loaded for the language `#1'. Using the pattern for}%
\typeout{** the default language instead.}%
\else
\language=\csname l@#1\endcsname
\fi
#2}}
\providecommand{\BIBdecl}{\relax}
\BIBdecl

\bibitem{farwell2011stuxnet}
J.~P. Farwell and R.~Rohozinski, ``Stuxnet and the future of cyber war,''
  \emph{Survival}, vol.~53, no.~1, pp. 23--40, 2011.

\bibitem{SRID_ESORICS14}
Y.~Wang, Z.~Xu, J.~Zhang, L.~Xu, H.~Wang, and G.~Gu, ``Srid: State relation
  based intrusion detection for false data injection attacks in scada,'' in
  \emph{Proceedings of the 19th European Symposium on Research in Computer
  Security (ESORICS'14)}, September 2014.

\bibitem{darpaCFP}
DARPA, ``Rapid attack detection, isolation and characterization systems
  (radics),''
  \url{http://www.darpa.mil/program/rapid-attack-detection-isolation-and-characterization-systems}.

\bibitem{antfarm}
S.~N. Lab, ``Advanced network toolkit for assessments and remote mapping,''
  \url{https://github.com/ccss-sandia/antfarm}.

\bibitem{smartcisco}
Cisco-Systems, ``Safe mapping and report tool,''
  \url{http://safemap.sourceforge.net/}.

\bibitem{netapt}
U.~of~Illinois~at Urbana-Champaign, ``Network access policy tool for
  verification of distributed and layered security policy implementation,''
  \url{https://www.perform.illinois.edu/netapt/}.

\bibitem{moteff2004critical}
J.~Moteff and P.~Parfomak, ``Critical infrastructure and key assets: definition
  and identification.''\hskip 1em plus 0.5em minus 0.4em\relax DTIC Document,
  2004.

\bibitem{franccois2011ptf}
J.~Fran{\c{c}}ois, H.~Abdelnur, R.~State, and O.~Festor, ``Ptf: Passive
  temporal fingerprinting,'' in \emph{Integrated Network Management (IM), 2011
  IFIP/IEEE International Symposium on}.\hskip 1em plus 0.5em minus 0.4em\relax
  IEEE, 2011, pp. 289--296.

\bibitem{gordeychik2013scada}
S.~Gordeychik, ``Scada strangelove or: How i learned to start worrying and love
  nuclear plants,''
  \url{https://events.ccc.de/congress/2012/Fahrplan/events/5059.en.html}, 2013.

\bibitem{matherly2009shodan}
J.~C. Matherly, ``Shodan the computer search engine,'' \emph{Available at
  [Online]: http://www. shodanhq. com/help}, 2009.

\bibitem{shine2014}
J.~B. Bob~Radvanovsky, ``Project shine: What we discovered and why you should
  care,'' 2015.

\bibitem{kiravuo2015peeking}
T.~Kiravuo, S.~Tiilikainen, M.~S{\"a}rel{\"a}, and J.~Manner, ``Peeking under
  the skirts of a nation: Finding ics vulnerabilities in the critical digital
  infrastructure,'' in \emph{Proceedings of the 14th European Conference on
  Cyber Warfare and Security 2015: ECCWS 2015}.\hskip 1em plus 0.5em minus
  0.4em\relax Academic Conferences Limited, 2015, p. 137.

\bibitem{bodenheim2014evaluation}
R.~Bodenheim, J.~Butts, S.~Dunlap, and B.~Mullins, ``Evaluation of the ability
  of the shodan search engine to identify internet-facing industrial control
  devices,'' \emph{International Journal of Critical Infrastructure
  Protection}, vol.~7, no.~2, pp. 114--123, 2014.

\bibitem{caselli2013feasibility}
M.~Caselli, D.~Had{\v{z}}iosmanovi{\'c}, E.~Zambon, and F.~Kargl, ``On the
  feasibility of device fingerprinting in industrial control systems,'' in
  \emph{Critical information infrastructures security}.\hskip 1em plus 0.5em
  minus 0.4em\relax Springer, 2013, pp. 155--166.

\bibitem{lin2008automatic}
Z.~Lin, X.~Jiang, D.~Xu, and X.~Zhang, ``Automatic protocol format reverse
  engineering through context-aware monitored execution.'' in \emph{NDSS},
  vol.~8, 2008, pp. 1--15.

\bibitem{clarke2004practical}
G.~R. Clarke, D.~Reynders, and E.~Wright, \emph{Practical modern SCADA
  protocols: DNP3, 60870.5 and related systems}.\hskip 1em plus 0.5em minus
  0.4em\relax Newnes, 2004.

\bibitem{jeon2014obfuscation}
S.~Jeon, J.-H. Yun, and W.-N. Kim, ``Obfuscation of critical infrastructure
  network traffic using fake communication,'' in \emph{Critical Information
  Infrastructures Security}.\hskip 1em plus 0.5em minus 0.4em\relax Springer,
  2014, pp. 268--274.

\bibitem{okhravi2014finding}
H.~Okhravi, T.~Hobson, D.~Bigelow, and W.~Streilein, ``Finding focus in the
  blur of moving-target techniques,'' \emph{Security \& Privacy, IEEE},
  vol.~12, no.~2, pp. 16--26, 2014.

\bibitem{ali2015randomization}
M.~Q. Ali and E.~Al-Shaer, ``Randomization-based intrusion detection system for
  advanced metering infrastructure,'' \emph{ACM Transactions on Information and
  System Security (TISSEC)}, vol.~18, no.~2, p.~7, 2015.

\end{thebibliography}

\end{document}